\documentclass[twocolumn,prl,aps,superscriptaddress,showpacs]{revtex4-1}

\usepackage{mathptmx}
\usepackage{dcolumn}
\usepackage{amsmath,amssymb}
\usepackage{bm}
\usepackage{color}
\usepackage{latexsym}
\usepackage{epstopdf}
\usepackage{color}
\usepackage[english]{babel}
\usepackage{latexsym}

\usepackage[pdftex]{graphicx}
\usepackage{amsmath} 
\usepackage{amssymb} 
\usepackage{amsfonts}
\usepackage{bm}
\usepackage{natbib}
\usepackage{epstopdf}
\usepackage{appendix}

\definecolor{mygrey}{gray}{0.35}
\definecolor{myblue}{rgb}{0.2,0.2,0.8}
\definecolor{myzard}{cmyk}{0,0,0.05,0}
\definecolor{mywhite}{rgb}{1,1,1}
\definecolor{mywhite}{rgb}{1,1,1}
\definecolor{myred}{rgb}{1,0.,0.3}

\usepackage[colorlinks=true,citecolor=myblue,linkcolor=myred]{hyperref}

\def\be{\begin{equation}}
\def\ee{\end{equation}}
\def\ba{\begin{align}}
\def\enda{\end{align}}
\def\bi{\begin{itemize}}
\def\ei{\end{itemize}}

\def\N{{\mathbb{N}}}

\def\dd{\mathord{\rm d}} 
 \def\ee{\mathord{\rm e}}
 
 \def\ii{\mathord{\rm i}}

\def\N{{\mathbb{N}}}

\def\dd{\mathord{\rm d}} 
 \def\ee{\mathord{\rm e}}
 
 \def\ii{\mathord{\rm i}}

\renewcommand{\ii}{{\rm i}}
\renewcommand{\ee}{{\rm e}}
\renewcommand{\aa}{{\rm a}}
\renewcommand{\dd}{{\rm d}}

\def\beq{\begin{equation}}
\def\beq{\begin{equation}}
\def\eeq{\end{equation}}

 \newcommand{\ket}[1]{|#1\rangle}
 \newcommand{\bra}[1]{\langle #1|}

\def\mmax{m_{\mathrm{max}}}
\def\oned{\mathrm{1D}}

\def\Nm{N_m}

\def\II{{\rm I}}

\def\cc{{\rm c}}
\def\dd{{\rm d}}

\def\ii{{\bf i}}

\def\DD{{\rm D}}

\def\PP{\mathbb{P}}
\def\QQ{\mathbb{Q}}

\newcommand{\ud}[1]{{#1^{\dagger}}}
\newcommand{\braket}[2]{\langle #1|#2\rangle}

\begin{document}

\title{Deterministic generation of arbitrary photonic states assisted by dissipation.}

 \author{A. Gonz\'{a}lez-Tudela}
 \affiliation{Max-Planck-Institut f\"{u}r Quantenoptik Hans-Kopfermann-Str. 1.
85748 Garching, Germany }

 \author{V. Paulisch}
 \affiliation{Max-Planck-Institut f\"{u}r Quantenoptik Hans-Kopfermann-Str. 1.
85748 Garching, Germany }
      
\author{D. E. Chang}
 \affiliation{ICFO-Institut de Ciencies Fotoniques, Mediterranean Technology Park, 08860 Castelldefels (Barcelona), Spain}

 \author{H. J. Kimble}
  \affiliation{Max-Planck-Institut f\"{u}r Quantenoptik Hans-Kopfermann-Str. 1.
85748 Garching, Germany }
 \affiliation{Norman Bridge Laboratory of Physics 12-33}
  \affiliation{Institute for Quantum Information and Matter, California Institute of Technology, Pasadena, CA 91125, USA}
  
\author{J. I. Cirac}
 \affiliation{Max-Planck-Institut f\"{u}r Quantenoptik Hans-Kopfermann-Str. 1.
85748 Garching, Germany }

\date{\today}

\begin{abstract}
A scheme to utilize atom-like emitters coupled to nanophotonic waveguides is proposed for the generation of many-body entangled states and for the reversible mapping of these states of matter to photonic states of an optical pulse in the waveguide. Our protocol makes use of decoherence-free subspaces (DFS) for the atomic emitters with coherent evolution within the
DFS enforced by strong dissipative coupling to the waveguide.  By switching from subradiant to superradiant states, entangled atomic states are mapped to photonic states with high fidelity. An implementation using ultracold atoms coupled to a photonic crystal waveguide is discussed.

\end{abstract}

\maketitle

\emph{Introduction.}
Recent experimental and theoretical work on optical emitters coupled to one-dimensional waveguides has opened new avenues to investigate light-matter interactions \cite{vetsch10a,goban12a,mitsch14a,petersen14a,mlynek14a,goban13a,yu14a,thompson13a,tiecke14a,goban15a,lodahl13a,arcari14a,beguin14a,sollner14a,goban15a,dzsotjan10a,gonzaleztudela11a,koppens11a,huidobro12a,martinmoreno15a}. Particularly promising are the setups where atoms are strongly coupled to structured dielectrics \cite{goban13a,yu14a,thompson13a,tiecke14a,goban15a}, where large Purcell factors have been predicted \cite{hung13a,douglas15a}. Furthermore, collective effects can be enhanced by placing the atoms at particular positions \cite{lehmberg70a,lehmberg70b,kien05a,dzsotjan10a,gonzaleztudela11a,chang12a,gonzaleztudela13a,shahmoon13a,zheng13a}. The combination of atom-like emitters and nanophotonic waveguides may enable new regimes for the interaction of light and matter, leading to technologies that outperform current ones and 
qualitatively different physics. In this work we investigate 
the possibility of using atom nanophotonics interfaces to tailor arbitrary states for propagating photons on demand. We predict large fidelities even for relatively large numbers of photons, something which has been impossible to achieve with other platforms in the optical domain.

\begin{figure}[tb]
\begin{center}
\includegraphics[width=0.42\textwidth]{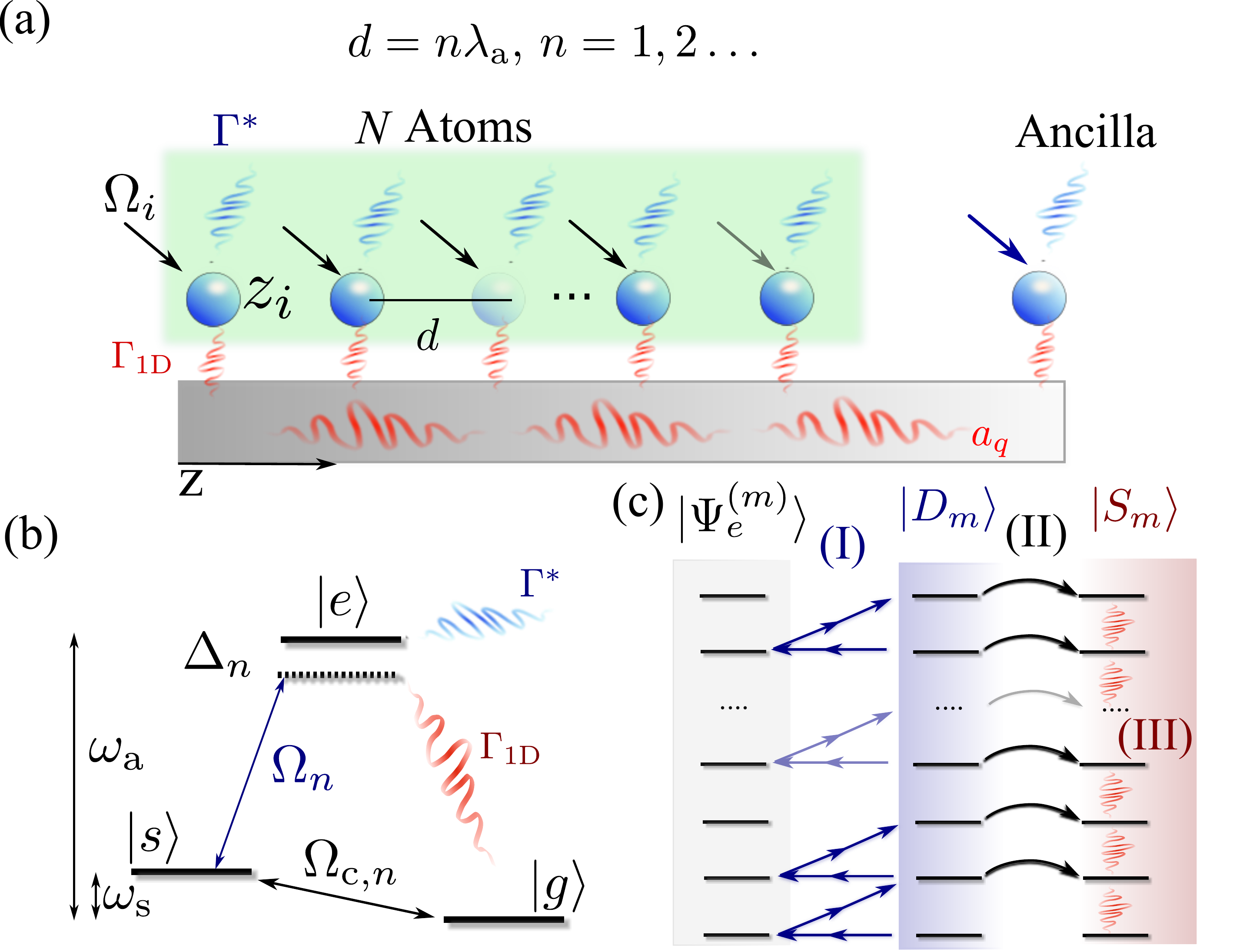}
\end{center}
\caption{(a) Scheme of the proposal: $N$ atoms equally spaced, $z_n=n\lambda_\aa=n 2\pi/q_\aa$, with $n\in \N$, plus $1$ ancilla atom coupled to a 1D photonic bath that can be separated by a larger distance as long as it satisfies $z_n=n\lambda_\aa$. The ancilla atom must be individually addressed. (b) Atomic $\Lambda$-scheme considered: the dipole transition $\ket{g}\leftrightarrow\ket{e}$ is coupled to the reservoir modes, $a_q$. A Raman laser controls the transition $\ket{s}\leftrightarrow\ket{e}$ with intensity, $\Omega_n$ and detuning $\Delta_n$. A second field controls $\ket{s}\leftrightarrow\ket{g}$, with frequency $\omega_s$. (c) Relevant states of the atomic ensemble for our three step protocol: (I) generation of superpositions of symmetric states, $\ket{D_m}$, in the two metastable states $\ket{g}$ and $\ket{s}$ by using the excited dark states $\ket{\Psi_e^{(m)}}$. (II) Flipping $\ket{s} \rightarrow \ket{e}$ to generate the superradiant state $\ket{S_m}$, which decays rapidly (III) to the desired 
entangled 
photonic state.}
\label{fig1}
\end{figure}

Our proposal uses three-level systems (with levels $\{\ket{g},\ket{s},\ket{e}\}$), where one of the optical transitions ($\ket{g}\leftrightarrow\ket{e}$) is strongly coupled to a one dimensional (1D) waveguide (see Fig.\ref{fig1}(a-b)). We denote by $P_{\oned}$ the Purcell factor corresponding to that transition; i.e., the ratio of the emission rate into the particular waveguide mode, $\Gamma_{\oned}$, and the one for all other modes, $\Gamma^*$. The atoms will be separated by distances proportional to the optical wavelength $\lambda_\mathrm{a} = 2 \pi/q(\omega_\mathrm{a})$, where $q(\omega)$ is determined by the waveguide dispersion relation. Depending on their internal state, atoms may experience a collective decay into the waveguide, or become completely decoupled from it. The latter occurs if they are in a decoherence free subspace (DFS)\cite{zanardi97a,lidar98a}, that is, if they are the states from the nullspace of the collective decay of the transition $\ket{e}\rightarrow\ket{g}$. Our protocol 
consists 
of two steps: in the first one, we generate certain decoherence-free states, $\ket{\Psi_D}$, by driving the atoms with lasers and using the collective quantum Zeno effect \cite{beige00a,facchi02a,law96a} within the DFS with a fidelity $1-F_1 \propto m/\sqrt{P_{\oned}}$, where $m$ is the maximum number of photons we want to generate; in the 
second one, a laser pulse takes the atomic state out of the DFS so that atoms collectively emit into the waveguide, creating the desired state of a single propagating mode, $\ket{\Psi_B}$. As we show, the second process has an (in)fidelity $1-F_2 \propto m^2/N P_{\oned}$, where $N$ is the number of atoms.

\emph{Coupling to a common environment}
The atom-photon hamiltonian of these systems is given by $H = H_0 + H_\II$, where $H_0= H_{\rm qb} + H_{\rm field}$ is the free energy term, with
$H_{\rm qb} =\sum_{n=1}^{N+1} ( \omega_\aa \sigma^{n}_{ee} + \omega_\mathrm{s} \sigma^{n}_{ss}$) and
$H_{\rm field} = \sum_q \omega_q a^\dagger_q a_q$,
 (using $\hbar = 1$) where $\omega_\aa$ is the two-level system energy, $\omega_q$ is the field dispersion relation of the 1D photonic modes, and $\sigma_{ij}^n=\ket{i}_n\bra{j}$. We consider only the coupling to a single polarization as justified for suitable dielectric waveguide modes \cite{hung13a}, that is, a dipolar coupling of the form
\begin{equation}
H_\II  = \sum_{n,q} \left( g_q \sigma_{ge}^n a^\dagger_q e^{-i q z_n} + \rm {H.c.} \right),
\label{Hint}
\end{equation}
with $g_q$ the single-photon coupling constant to the mode of interest and where we have used the rotating wave approximation. Under the condition that the 1D-bath degrees of freedom have a much faster relaxation timescale than for the atomic system, the atoms are described by a density matrix, $\rho$, which in the Born-Markov limit, is governed by a master equation  \cite{gardiner_book00a,lehmberg70a,lehmberg70b} of the form: $d \rho /d t = \sum_{n,m} \frac{\Gamma_{\oned}}{2} e^{i q(\omega_\aa) |z_n - z_m |} \left( \sigma_{ge}^n \rho \sigma_{eg}^m - \rho \sigma_{eg}^m \sigma_{ge}^n \right)
+ \rm {H.c.}$, where $\Gamma_{\oned}$ is the renormalized spontaneous emission rate due to the interaction with the 1D photonic reservoir. 

\emph{Emergence of subradiant and superradiant states.}
By appropriately choosing the atomic positions, e.g., $z_n = n \lambda_\aa=n 2\pi/q(\omega_\aa)$, with $n\in \N$, the coherent atom-atom interactions are eliminated \cite{footnote1} and the effective interaction yields a pure Dicke model \cite{dicke54a} decay described by 
\begin{equation}
{\cal L}_{\rm D}(\rho) = 
\frac{\Gamma_{\oned}}{2} \left(S_{ge} \rho S_{eg} -  S_{eg} S_{ge} \rho \right) + \rm H.c.,
\label{eq:Dicke}
\end{equation}
where we have introduced the collective spin operators $S_{ij} = \sum_{n=1}^{N+1} \sigma^n_{ij}$. One of the assets of the model described by Eq.~\ref{eq:Dicke} is the emergence of \emph{sub} and \emph{super}radiant states. The excited states with $m$ atoms in $\ket{e}$ that are symmetric under the permutation of atoms, denoted by $\ket{S_m}$, are superradiant with an enhanced decay rate proportional (at least) to the atom number $N$, and are unique for each $m$. On the other hand, the states satisfying $S_{ge}\ket{\Psi}=0$ are dark states of the Liouvillian of Eq.~\ref{eq:Dicke}, and are therefore decoupled from collective dissipation. These dark states span the DFS that is highly degenerate for $m>1$.

The atomic entangled states that have to be created in the first step of our protocol are very peculiar, as: i) they have to belong to the DFS; ii) they must be prepared within that subspace to avoid dissipation; iii) and they must have a special form such that they can be easily mapped in the second step to the appropriate superradiant states, in order to generate arbitrary superpositions of the photonic states in the waveguide. Our strategy consists of first, identifying states, denoted by $\ket{D_m}$, in the subspace spanned by the ground levels $g$, and $s$, which can be mapped one-to-one to a basis $\ket{S_m}$ of superradiant states using a simple laser pulse, and which in turn give rise to $m$ photons in the waveguide via superradiance. Then, we use a more sophisticated scheme within the DFS to generate superpositions of $\ket{D_m}$, which requires only $m$ steps. 

By introducing an auxiliary metastable state, $\ket{s}$, as depicted in Fig.\ref{fig1}(b), the candidate states to map to superradiant ones are the symmetric Dicke states $\ket{D_m}\propto \mathrm{sym} \{\ket{s}^{\otimes m}\otimes \ket{g}^{\otimes N-m}\}$, as they can be turned superradiant by switching $\ket{s}\rightarrow\ket{e}$. Having identified the target, $\ket{D_m}$, we need to find ways to build efficiently arbitrary superpositions. One general strategy already explored in the literature consists of implementing one and two-qubit universal gates within the DFS \cite{zanardi97a,lidar98a,beige00a,facchi02a}, which in principle allows to build any arbitrary superposition. However, the number of steps increases rapidly with the number of atoms. Another possibility is to use superposition  of Zeeman states of $1$ atom and rapid adiabatic passage methods  \cite{parkins93a,parkins95a,lange00a}, but it is limited by the number of metastable states available. For small system sizes other adiabatic passage methods have been numerically studied \cite{shao10a}; however, they do not work for superposition states. In our proposal, we use the collective character of the interaction to deterministically generate arbitrary $N$-qubit states for which the number of steps is independent of the number of atoms.

\emph{Controlling atomic states under strong dissipation.}

The scheme that we use is depicted in Fig.~\ref{fig1}(a): we consider a system of $N+1$ emitters, in which we aim to generate $\ket{D_m}$ in the first $N$ emitters using the ancilla as an auxiliary state. As the $\ket{D_m}$'s are invariant under the permutation of the first $N$ atoms, we choose the control fields with the same symmetry. They can be described by the following hamiltonian (in the interaction picture with respect to $H_\mathrm{qb}$):
\begin{align}
H_{\mathrm{c}}&=\frac{\Omega_{\mathrm{c}}}{2} \sigma^{N+1}_{sg} + \mathrm{H.c.}\, , 
\label{eq:Ham} \\
H_{\mathrm{las}}&=\left( \frac{\Omega_\mathrm{r}}{2}\sum_{n=1}^{N} \sigma^n_{se} + \frac{\Omega_\mathrm{anc} }{2} \sigma^{N+1}_{se}+\mathrm{H.c.} \right) + \Delta_e \sum_{n=1}^{N+1} \sigma_{ee}^n\,, \nonumber
\end{align}
where $H_{\mathrm{las}}$ allows to control both the emitter state and the coupling to the 1D-reservoir, while $H_{\mathrm{c}}$ allows to control the atomic states of the ancilla independently of the coupling to the 1D-reservoir. We are interested in working in the regime of strong collective dissipation, where 
$ N \Gamma_{\oned}\gg\Omega_\mathrm{r},\Omega_\mathrm{anc},\Omega_\mathrm{c},\Delta_e$.
In this situation, the 1D-bath is continuously monitoring the collective atomic state, as in the Quantum Zeno regime \cite{zanardi97a,lidar98a,beige00a,facchi02a}, and projecting the atomic state into the DFS of the Liouvillian $\mathcal{L}_\mathrm{D}$. 
Formally, we obtain the effective dynamics within the DFS by treating the control fields and dissipation into other decay channels as a perturbation to the collective dissipation $\mathcal{L}_\DD$ \cite{supmat}. For the ideal protocol, we consider the first order, that is the 
action of the control fields projected 
into the DFS. The errors due to higher order terms are discussed later.

\emph{Preparation of many-body entangled states: ideal protocol.}
Due to the symmetry of the problem, it is convenient to introduce the following notation to describe any symmetric state over $N$ atoms:
\begin{align}
\ket{F_{m,k}}\propto \mathrm{sym}\{\ket{s}^{\otimes m}\otimes\ket{e}^{\otimes k}\otimes\ket{g}^{\otimes N-m-k}\}\,, 
\end{align}
that embeds both $\ket{D_m}\equiv \ket{F_{m,0}}$ and  $\ket{S_m}\equiv \ket{F_{0,m}}$. In general, the DFS of the Liouvillian $\mathcal{L}_\mathrm{D}$ grows exponentially with the number of atoms, but for each $m$, only three of these states fulfill the permutation symmetry of our system \cite{supmat}. These states are:
\begin{align}
\ket{\Psi_{s}^{(m)}}=& \ket{F_{m-1,0}}\otimes\ket{s}_\mathrm{A}\,,\, \ket{\Psi_{g}^{(m)}}=\ket{F_{m,0}}\otimes\ket{g}_\mathrm{A}\,, \\
\ket{\Psi_e^{(m)}}=&\sqrt{\frac{\Nm}{\Nm+1}}\ket{F_{m-1,0}}\otimes\ket{e}_\mathrm{A} -\frac{1}{\sqrt{\Nm+1}}\ket{F_{m-1,1}}\otimes\ket{g}_\mathrm{A}, \nonumber
\end{align}
where $\ket{\psi}_\mathrm{A}$ denotes the state of the ancilla atom and $\Nm = N-m+1$. In Fig.~\ref{fig2} we sketch the protocol steps by a diagram of the projected hamiltonian into the DFS. It consists of two parts, which are applied $\mmax$ times to reach any superposition of states over the first $N$ atoms $\ket{\Psi_D}\otimes \ket{g}_\mathrm{A} = (\sum_{m=0}^{\mmax} d_m \ket{D_m})\otimes \ket{g}_\mathrm{A}$ with maximum $\mmax$ excitations from the initial state $\ket{\Psi_g^{(0)}}$:

\begin{figure}[tb]
\begin{center}
\includegraphics[width=0.4\textwidth]{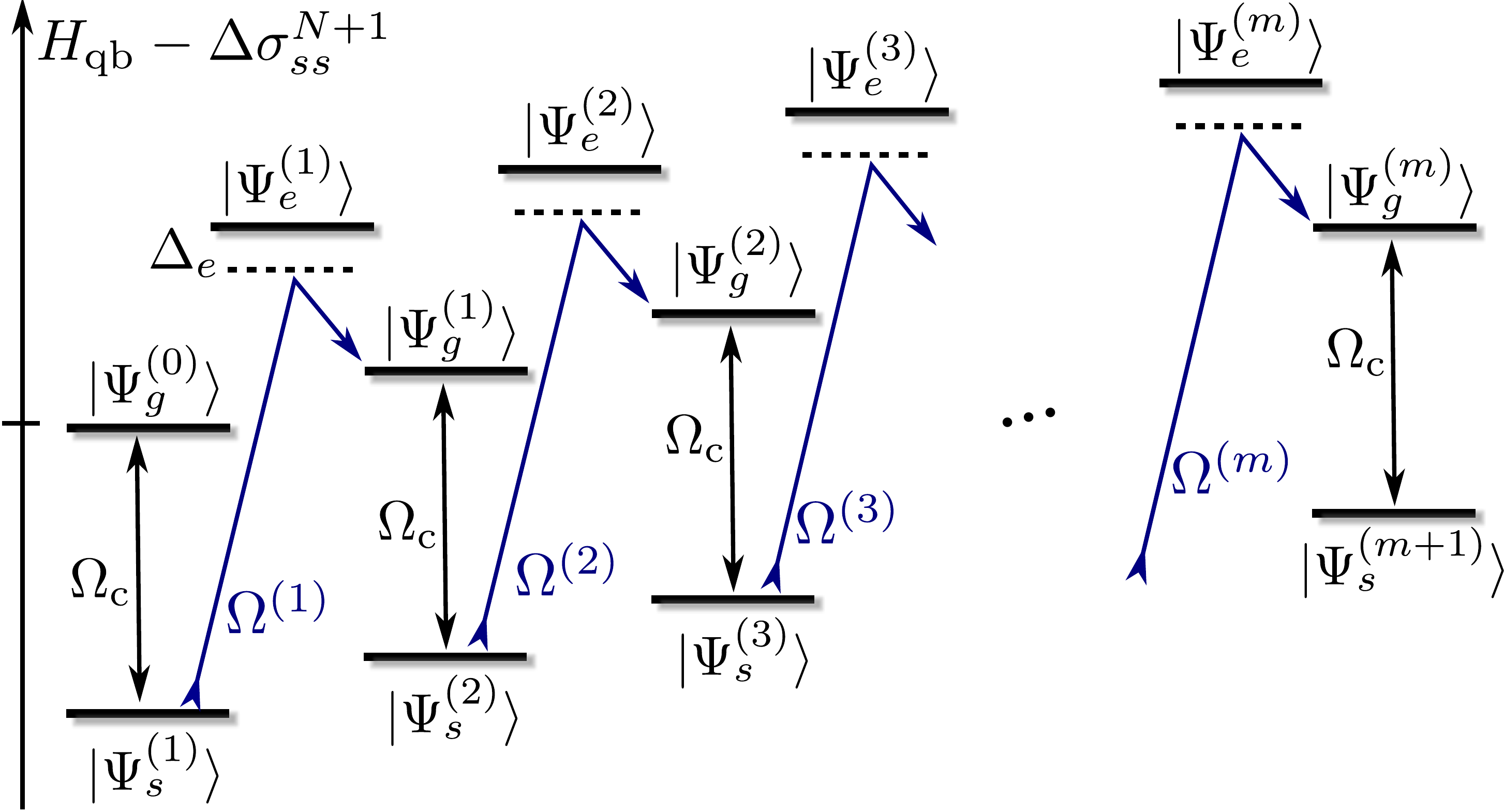}
\end{center}
\caption{Scheme for the preparation of arbitrary superpositions of symmetric Dicke states $\ket{D_m}$ of $N$ atoms. Alternating between $\sigma_x$ gates on the ancilla and two-photon transitions via $\ket{\Psi_e^{(m)}}$ builds up excitations step by step (from left to right). To make the ladder-like structure clearer, we added a term for the ancilla in the energy scale. }
\label{fig2}
\end{figure}

\begin{enumerate}
\item Use $H_\cc$ to flip the ancilla state $\ket{g}_\mathrm{A} \rightarrow \ket{s}_\mathrm{A}$. This transition makes: $\ket{\Psi_g^{(m-1)}} \rightarrow \ket{\Psi_s^{(m)}}$.

\item Two-photon transition $\ket{\Psi_s^{(m)}} \rightarrow \ket{\Psi_g^{(m)}}$. It can be shown that the dark states corresponding to a given excitation $m$ form an effective $\Lambda$-scheme within the DFS via the far detuned state $\ket{\Psi_e^{(m)}}$ \cite{supmat}. The two-photon transition can be made on-resonance if the Rabi couplings are chosen such that $|\Omega_{\mathrm{anc}}|=|\Omega_\mathrm{r}|\sqrt{\frac{m}{\Nm}}$, which is possible because we demanded individual addressing of the ancilla. Notice that if this condition is not imposed, the $\ket{\Psi_{s,g}^{(m)}}$ experience different Stark-shifts that ultimately spoil the two-photon process. The effective Hamiltonian for the $m$-th excitation is
\begin{align}
H_\DD = \Omega^{(m)}/2 \ket{\Psi_{s}^{(m)}} \bra{\Psi_{g}^{(m)}} + \rm H.c.,
\end{align}
where $|\Omega^{(m)} |=\frac{|\Omega_\mathrm{r}|^2}{2|\Delta_e|} \frac{m}{\Nm+1}$ and where the Stark shifts compensate each other and can thus be neglected.
\end{enumerate}

As depicted in Fig.~\ref{fig2}, the combination of both steps gives rise to a ladder-like structure, which can be used to build any arbitrary superpositions state of $\ket{D_m}$ from the ground state $\ket{\Psi_g^{(0)}}$. One way of finding the pulse sequence to create any arbitrary superposition consists of removing the excitations in the system, starting from $\ket{\Psi_g^{(\mmax)}}$ step by step until only $\ket{\Psi_g^{(0)}}$ is populated \cite{law96a}.  Remarkably, the number of steps required depends only linearly on the number of excitations $\mmax$, independently of the number of atoms, in stark contrast to building the states out of one and two-qubit gates \cite{beige00a}.

\emph{Shaping an arbitrary field of light using strong dissipation.}
Apart from the intrinsic interest of generating arbitrary superpositions $\ket{\Psi_{D}}$, another advantage is that our protocol can be used to generate a superposition of bosonic states in the 1D-bath by dissipative means. To make sure the ancilla atom can be neglected, we flip the ancilla state $\ket{g}_\mathrm{A} \rightarrow \ket{s}_\mathrm{A}$ and apply no fields to it. In order to map to the superradiant state of $N$ atoms, we apply a fast resonant $\pi$-pulse ($\Delta_e=0$ and $\Omega_\mathrm{r} \gg N\Gamma_{\oned}$) on the $N$ emitters to switch all $\ket{s}_n\rightarrow \ket{e}_n$, thus generating the superposition of $\ket{S_m} \propto \mathrm{sym}\{\ket{e}^{\otimes m}\otimes\ket{g}^{\otimes N-m}\}$. Due to their superradiant character, the  $\ket{S_m}$ decay completely to 1D-reservoir modes. Because $H_\II$ conserves the number of excitations, the superradiant state of $m$-excitations, decays to Fock-state of $m$-photons \cite{supmat}:
\begin{equation}
\ket{S_{m}}\rightarrow \ket{m_{\{q\}}}\equiv \sum_{\{q\}} \frac{A_{\{q\}}(t)}{m!}a_{q_1}^\dagger \dots a_{q_m}^\dagger\ket{\mathrm{vac}}\,,
\end{equation}
where $\{q\}=\{q_1,\dots,q_m\}$ is the set of relevant momenta of the $m$-photon Fock state, where each of them run over the whole Brillouin Zone $q_i\in\mathrm{B.Z.}$. The scattering amplitude $A_{\{q\}}(t)$ can be calculated using a generalized input-output formalism \cite{caneva15a,xu15a,supmat} and quantum regression theorem \cite{lax63a}:
\begin{align}
		A_{ \{ q \} }(t) =  \prod_{r=1}^{m} \frac{\ii g \sqrt{r N_r}\ \ee^{-\ii \omega_{q_r} t} }{ \ii ( \sum_{l=1}^r \omega_{q_l}-r \omega_\aa) + r \Gamma_\oned N_r/2} +\mathrm{P}[\{q\}]\,
		\label{eq:amplitude}
\end{align}
for sufficiently large times $t \gg 1/ N_m \Gamma_\oned$ when the atomic state has decayed completely and defining $\mathrm{P}[\{q\}]$ as all the permutations of $\{q\}$. Notice that the only dependence on $t$ in this case enters through: $\ee^{-\ii \sum_{r=1}^m \omega_{q_r} t} $, which describes the center of mass motion of the wavepacket when going to the real space. In the low excitation regime, one can either use the Holstein-Primakoff approximation \cite{porras08a} or directly substitute $N_m\rightarrow N$ in the expression above. In both cases, we arrive to \cite{supmat}:
\begin{align}
		A^{\mathrm{HP}}_{ \{ q \} }(t) =  \prod_{r=1}^{m} \frac{\ee^{-\ii \omega_{q_r} t}\ii g \sqrt{ r N} }{ \ii ( \omega_{q_r}- \omega_\aa) + \Gamma_\oned N/2}\,,\label{eq:photonHP}
\end{align}
that has a Lorentzian shape centered at $\omega_\aa$ with bandwidth $\Gamma_{\oned} N/2$. Substituting $A_{ \{ q \} }(t)\rightarrow A^{\mathrm{HP}}_{ \{ q \} }(t)$ into the definition of $\ket{m_{\{q\}}}$, yields a linear Fock state denoted by $\ket{m^{\mathrm{HP}}_{\{q\}}}$. In principle, the emission into the waveguide is bidirectional ($\pm q$), but combining both fields in phase, e.g., by a placing a mirror at an appropiate distance, or by engineering the atom-photon coupling appropriately \cite{mitsch14a,petersen14a,sollner14a}, it is possible to achieve emission in one-direction only. Furthermore, by shaping the pulse, $\Omega_\mathrm{r} (t)$ (within a bandwidth $\lesssim N\Gamma_{\oned}$) we generate any desired shape of the output photonic state, e.g., to create a time-symmetric photonic state \cite{cirac97b} that ensures the reversibility of the process when mapping the photonic state to another sample. Moreover, because of the linearity of the calculation of $A_{\{q\}}$ with respect to the input 
state \cite{supmat}, 
superpositions of atomic states decay to superpositions:
\begin{equation}
\ket{\Psi_{D}}\rightarrow \sum_{m=0}^{m_{\mathrm{max}}} d_m \ket{S_{m}} \rightarrow \ket{\Psi_\mathrm{B}}=\sum_{m=0}^{m_{\mathrm{max}}} d_m \ket{m_{\{q\}}}\,,
\end{equation}
that will be generated in a single-mode wavepacket as long as $N\gg m$ because: $\braket{m_{\{q\}}}{m^{\mathrm{HP}}_{\{q\}}}=1-m^3/(20 N^2)+O(m^4/N^3)$ \cite{supmat}.

\emph{Preparation of many-body entangled states: error analysis.}
So far, we have only discussed the ideal protocol without considering detrimental effects, e.g., spontaneous emission into all other modes with rate $\Gamma^*$.  For the error in the preparation of the many-body entangled state, we derive an error rate $\epsilon$ from perturbation theory, which, together with the time of the operation $\tau$, gives an approximation of the infidelity of the state, $1-F_1 \approx \tau \epsilon$, where the fidelity $F_1$ is calculated with respect to the target state $\ket{\Psi_\DD}$, i.e., $F_1 =\sqrt{\bra{\Psi_\DD} \rho(\tau) \ket{\Psi_\DD} }$.

The dominant errors assuming $N\gg m$ and $\Gamma_{\oned}\gg\Delta_e\gg \Gamma^*$ (see Ref. \cite{supmat} for complete expressions) come from:  i) the spontaneously emitted photons from $\ket{\Psi_e^{(m)}}$ to decay channels other than the waveguide. As we are using an off-resonant Raman transition, this error scales as $\epsilon_1 \approx  \Gamma^*\frac{m|\Omega_r|^2}{4N\Delta_{e}^2}$. ii) Other errors appear due to the photons emitted from the small populations of superradiant states. We estimate the rate of these errors by taking into account the second order corrections of the projected hamiltonian \cite{supmat}, which are finally given by $\epsilon_2 \approx N \Gamma_\oned \frac{m |\Omega_r|^2}{4(\Delta_{e}^2+(N \Gamma_\oned)^2)}$. Summing up, the infidelity for the $m$-th step of the process, which takes $\tau =\frac{\pi}{|\Omega^{(m)}|} \approx  \frac{2\pi N \Delta_e}{m|\Omega_r|^2} $ for full population transfer, is
\begin{equation}
1-F_1^{(m)}\approx \frac{\pi}{2} \Big(\frac{\Gamma^*}{\Delta_e}+\frac{\Delta_e}{\Gamma_{\oned}}\Big),
\end{equation}
that is optimised for $\Delta_{e,\mathrm{opt}}=\sqrt{\Gamma^* \Gamma_{\oned}}$, which yields a scaling: $1-F^{(m)}_{1,\mathrm{opt}}\propto 1/\sqrt{P_{\oned}}$. Interestingly, the scaling depends neither on the number of atoms, $N$, nor the number of excitations, $m$.  In order to create any superposition $\ket{\Psi_{D}}$, we require $m_{\mathrm{max}}$ steps, thus, the total infidelity of the first part of the protocol is
\begin{equation}
1-F_{1}\propto \frac{m_{\mathrm{max}}}{\sqrt{P_{\oned}}}\,,
\label{eq:Fscaling}
\end{equation}
that can be improved via post-selection of the states condition on detecting no photons to the waveguide. \cite{footnote2}

\emph{Preparation of many-body entangled states: numerical analysis.} To validate the scaling analysis, we study numerically (without any approximation) the preparation of two relevant sets of states:  i) the general class of states $\ket{D_m}$; ii) the superpositions $\ket{\Phi_m}=\frac{1}{\sqrt{2}}\big(\ket{D_0}+\ket{D_m}\big)$. First, recall that the symmetry conditions that we found from our analysis of the ideal situation tell us that the relevant Hilbert space does not depend on the atom number, $N$, but only on the maximum number of excitations, $\mmax$ \cite{supmat}. The atom number only enters on the two-photon resonance condition, that fixes $\Omega^{(m)}$. Using this restricted Hilbert space, we use a non-hermitian evolution determined by the coherent driving plus the imaginary energies determined by the coupling to the waveguide through the $S_{ge}$ 
operator, to give $H_\mathrm{eff} = H_{\mathrm{las}}+H_{\cc} -\ii \Gamma_\oned S_{eg} S_{ge}/2  - \ii \Gamma^* S_{ee}/2$.

\begin{figure}[!t]
\begin{center}
\includegraphics[width=0.45\textwidth]{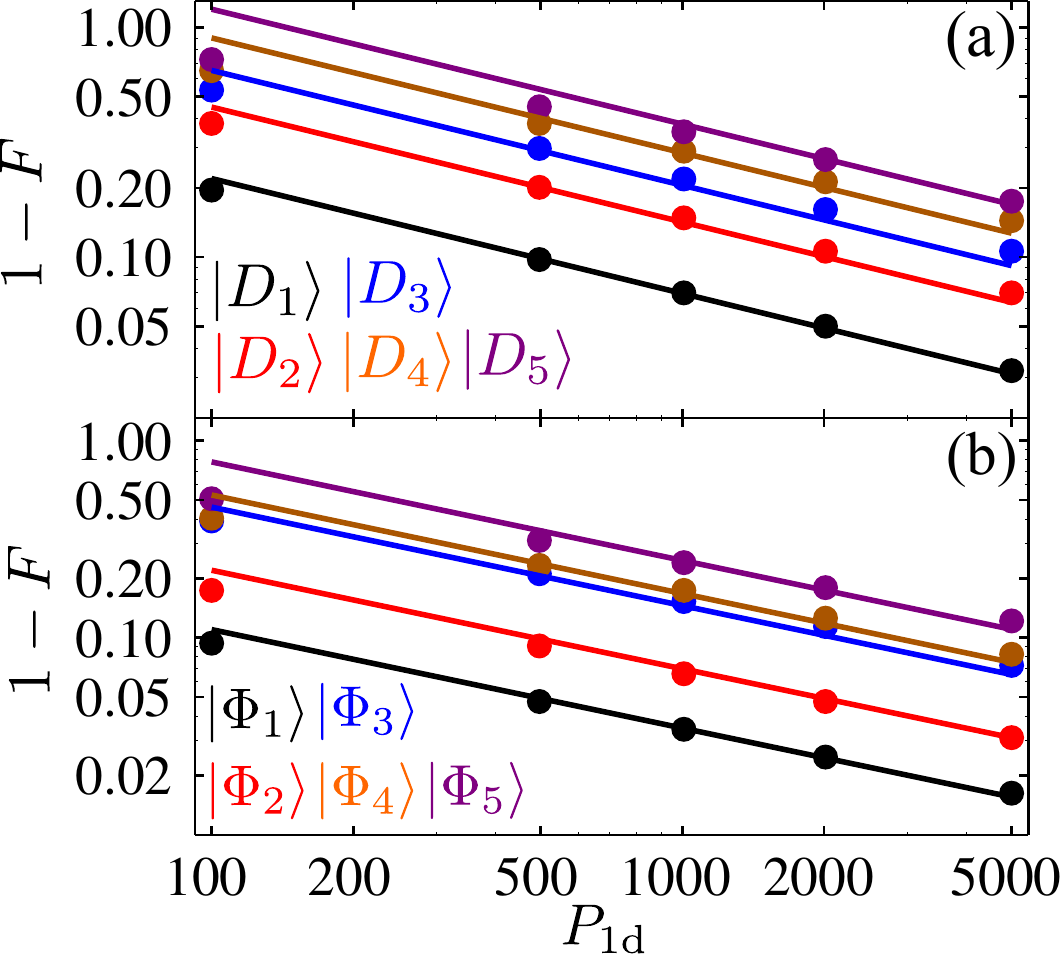}
\end{center}
\caption{(a) [(b)] Infidelities as a function of $P_\oned$ for generating $\ket{D_m}$ [$\ket{\Phi_m}=\frac{1}{\sqrt{2}}\big(\ket{D_0}+\ket{D_m}\big)$] up to $\mmax=5$ excitations. The dots correspond to the numerical fidelities, whereas the solid lines depict the $1/\sqrt{P_{\oned}}$ scaling from Eq.~\ref{eq:Fscaling}. }
\label{fig3}
\end{figure}

For the generation of Fock states, the pulse sequence consist of a complete transfer of populations in each step of Fig.~\ref{fig2}: 
$\ket{\Psi_g^{(0)}} \rightarrow \ket{\Psi_s^{(1)}} \rightarrow \dots \rightarrow \ket{\Psi_g^{(m)}}$, which can be ensured by fixing the time of interaction, $t$, to $t\Omega_{\mathrm{c}}=\pi$ ($t\Omega^{(m)}=\pi$) for the microwave (Raman) transitions. For a more complicated state, such as the $\ket{\Phi_m}$ the pulse sequence can be calculated numerically. In Fig.~\ref{fig3}(a-b), we show the numerical fidelities obtained when fixing the off-resonant transition to the optimal $\Delta_{e,\mathrm{opt}}$ that we found from our error analysis. The plots show how our general arguments give the correct scaling $\propto 1/\sqrt{P_{\oned}}$.

\emph{Generation of Photonic States: Error Analysis.}
The errors from deviations from the Holstein-Primakoff approximation, which leads to Eq.~\ref{eq:photonHP}, have already been discussed. Here, we take into account the emission into other modes through $\Gamma^*$ to obtain an estimate for the fidelity, that for a state with $m$-excitations is given by $m \Gamma^*$. Then, the infidelity of an operation with average time  $\frac{1}{N\Gamma_\oned}$ is approximately $1-F_2^{(m)}  =\frac{2 m \Gamma^*}{N \Gamma_{\oned}}$. When there are $\mmax$ excitations in the system, one obtains an infidelity for the mapping to the photonic state of the process:
 \begin{align}
 1-F_{2} \approx \frac{\mmax^2  \Gamma^*}{N \Gamma_\oned} = \frac{\mmax}{N} \frac{\mmax}{P_\oned}.
 \end{align}

Remarkably, the dissipative character of this mapping allows for the deterministic and efficient generation of (arbitrary superpositions of) photonic states, e.g., Fock states, that typically must be generated probabilistically \cite{dakna99a,afek10a} with an exponential scaling with the number of excitations or via non-linear interactions \cite{law96a}.
 
\emph{Implementation.}
In dielectric photonic crystal waveguides, $\Gamma_{\oned}/\Gamma_\aa= \xi n_g\sigma/(2A_m)$, where $n_g=c/v_g$ is the group index, $\sigma=3 \lambda_0^2/(2\pi)$ the radiative cross-section, $A_m$ the effective mode area, $\Gamma_\aa$ the vaccuum emission rate and $\xi$ an adimensional factor coming from cavity enhancement. Current values for Cs atoms ($\lambda_0=894$ nm, $\Gamma_\aa/2\pi= 5.02$ MHz) and SiN alligator waveguides \cite{goban13a,goban15a} have $A_m \approx 0.2$ $\mu$m$^2$, $n_g\approx 10$, $\xi\sim 5$ and $Q$-factors of $10^6$. Depending on the reduction of spontaneous emission, $\Gamma^*=\alpha\Gamma_\aa$, these numbers lead to $P_{\oned}\approx 50/\alpha$. Intrinsic losses in the dielectric (finite $Q$-factors) and reduced $v_g$ set finite propagation lengths of waveguide modes, $L_{\mathrm{prop}}/\lambda_\aa\approx Q/(2\pi n_g)$, which is $>10^4$ for state-of-art SiN values \cite{goban13a,goban15a}. Retardation effects also set a maximum number of atoms for which superradiant atom-photon 
mapping could be 
observed, e.g., assuming a separation $\lambda_\aa/2$, then $N\Gamma_{\oned}<2 v_g/(N \lambda_\aa)$, which for current structures leads to $N\lesssim 500$ atoms. Possible ways of overcoming retardation are to increase $\Gamma_{\oned}$ not only by $v_g$ but through cavity enhancement, e.g., by using mirrors \cite{goban15a}; or directly, by doing the atom-photon mapping off-resonantly, which decreases effectively $\Gamma_{\oned}$ (and $\Gamma^*$ by the same amount) relaxing the retardation requirements. Other potential problems that we have neglected include i) imperfect atomic separations limited ultimately by quantum center of mass wavepackets and atomic motion and ii) group velocity dispersion that distorts the wavepacket as they propagate through the waveguide. These effects should be studied in detail to evaluate and minimize the impact on the performance on the protocols described.

\emph{Conclusions.}
We have presented a protocol to generate deterministic superpositions of many-body entangled atomic states in the presence of strong dissipation. Remarkably, the infidelities in the preparation of complex superposition states increase only linearly with the number of excitations of the system ($m$), inversely with the square root of the Purcell Factor ($1/\sqrt{P_{\oned}}$), and are independent of the number of atoms considered. Furthermore, we have shown how to map these atomic states to photonic states with a very efficient scaling that depends linearly on the inverse collective Purcell Factor $1/(N P_{\oned})$ and how to engineer a time-symmetric wavepacket that guarantees the reversibility of the mapping.  A possible outlook of this work is to use other metastable states and different polarizations to generate entangled photonic NOON states in $\pm q$ directions \cite{petersen14a,mitsch14a,sollner14a} which is very relevant for quantum metrology applications \cite{giovannetti04a,afek10a}. 

\textbf{Acknowledgements} -
 We gratefully acknowledge discussions with T. Shi. The work of AGT, VP and JIC was funded by the European Union integrated project \emph{Simulators and Interfaces with Quantum Systems} (SIQS). AGT also acknowledges support from Alexander Von Humboldt Foundation and Intra-European Fellowship NanoQuIS (625955). JIC acknowledges support as a Moore Distinguished Scholar. DEC acknowledges support from Fundacio Privada Cellex Barcelona, Marie Curie CIG ATOMNANO, and ERC Starting Grant FoQAL. HJK acknowledges funding by the Institute of Quantum Information and Matter, a National Science Fundation (NSF) Physics Frontier Center with support of the Moore Foundation, by the Air Force Office of Scientific Research, Quantum Memories in Photon-Atomic-Solid State Systems (QuMPASS) Multidisciplinary University Research Initiative (MURI), by the  Department of Defense National Security Science and Engineering Faculty Fellows (DoD NSSEFF) program, by NSF PHY1205729 and support as a Max Planck Institute for Quantum Optics 
Distinguished Scholar.

\newpage

\widetext
\begin{center}
\textbf{\large Supplemental Material: Deterministic generation of many-body entanglement and photonic states assisted by dissipation.}
\end{center}
\setcounter{equation}{0}
\setcounter{figure}{0}
\makeatletter

\renewcommand{\thefigure}{SM\arabic{figure}}
\renewcommand{\thesection}{SM\arabic{section}}  
\renewcommand{\theequation}{SM\arabic{equation}}  

\section{Preparation of $\ket{D_{m}}$ . \label{appendixI} }

\subsection{Coupling to a common environment.}

Let us consider first the hamiltonian part describing the atom-photon coupling for an optical transition $\ket{g}\longleftrightarrow\ket{e}$, coupled to 1D photonic reservoir. In this case, the hamiltonian is given by $H = H_0 + H_\II$, where $H_0$ is the free term:
$H_0 = H_{\rm qb} + H_{\rm field}$, (using $\hbar = 1$)
\begin{equation}
H_{\rm qb} = \omega_\aa  \sum_{n=1}^{N+1} \sigma^{n}_{ee}, \ \ \ \
H_{\rm field} = \sum_q \omega_q a^\dagger_q a_q,
\label{eqS:H0}
\end{equation}
where $\omega_\aa$ is the atomic transition energy, $\omega_q$ is the field frequency from the dispersion relation of the waveguide modes. We consider that we are coupled to a single polarization to focus on the most relevant physics of our work which can be justified for appropriately designed dielectric waveguides \cite{hung13a}. We consider a dipolar coupling of the form
\begin{equation}
H_\II  = \sum_{n=1}^{N+1} \left( \sigma_{ge}^n E(z_n) + \rm {H.c.} \right),
\label{Hint}
\end{equation}
with $E(z) = \sum_q g_q a^\dagger_q e^{-i q z}$, and $g_q$ the single-photon coupling constant. In the conditions where the 1D-bath degrees of freedom have a much faster relaxation timescale than the system, we can describe our atomic system via a density matrix, $\rho$, which in Born-Markov limit, is governed by a master equation: $d \rho /d t = {\cal L_\DD}(\rho)$ \cite{gardiner_book00a,lehmberg70a,lehmberg70b}, with the superoperator
\begin{equation}
\label{eqS:mequation1}
{\cal L_\DD}(\rho) = 
\sum_{n,m} \Gamma_{n,m} \left( \sigma_{ge}^n \rho \sigma_{eg}^m - \rho \sigma_{eg}^m \sigma_{ge}^n \right)
+ \rm {H.c.} \,,
\end{equation}
where $\Gamma_{n,m} = \frac{\Gamma_{\oned}}{2} e^{i q(\omega_\aa) |z_n - z_m |}$, with $\Gamma_{\oned}$ the renormalized space spontaneous decay rate of the atoms due to the interaction with the 1D photonic reservoir. For completeness, it is interesting to write the connection between $\Gamma_{\oned}$ and $g_q$ that can be easily computed from \cite{gardiner_book00a}:
\begin{align}
\Gamma_{\oned}=2\pi\sum_q |g_q|^2\delta(\omega_\aa-\omega_q)=  L\int_{-\infty}^\infty \dd q |g_q|^2\delta(\omega_\aa-\omega_q)=\frac{2L |g_{q_\aa}|^2}{v_g(\omega_\aa)}\,, \label{eqS:gammaoned}
\end{align}  
where we have introduced $L$ the quantization length of the guided modes and used the fact that $\omega_q=\omega_{-q}$ and $|g_q|=|g_{-q}|$. 

\subsection{Emergence of subradiant and superradiant states.}

In the main manuscript we stipulated a homogeneous coupling to the environment, that can be achieved naturally with a 1D reservoir by choosing appropriately the atomic positions, i.e., $z_n = n \lambda_\aa=n 2\pi/q(\omega_\aa)$, with $n \in \N$, along the waveguide. With this choice, the effective interaction induced by reservoir modes yields a pure Dicke model \cite{dicke54a} decay described by 
\begin{equation}
{\cal L}_{\rm D}(\rho) = 
\frac{\Gamma_{\oned}}{2} \left(S_{ge} \rho S_{eg} -  S_{eg} S_{ge} \rho \right) + \rm H.c.,
\label{eqS:Dicke}
\end{equation}
where we have introduced the following notation for the collective spin operators $S_{ij} = \sum_{n=1}^{N+1} \sigma^n_{ij}$. $S_{eg}$ is just the collective operator for the spin dipole $\sigma_{eg}$. One of the assets of the model described by Eq. \ref{eqS:Dicke} is the emergence of  \emph{sub} and \emph{super}radiant states as depicted in Fig. \ref{fig1}(c), which can be seen easily by examining $S_{ge}\ket{J,m_J,\alpha_J}=\sqrt{(J+m_J)(J-m_J+1)}\ket{J,m_J-1,\alpha_J}$ in the collective angular momentum basis $\{\ket{J,m_J,\alpha_J}\}$ with $J=N/2,N/2-1,\dots$ and $m_J=-J,-J+1,\dots J$ and $\alpha_J$ is an index that takes into account the degeneracy of the states (that we drop from here on as it does not play an important role for what we will describe). The states satisfying $S_{ge}\ket{\Psi}=0$ (i.e., $m_J=-J$) are dark states of the Liovillian of Eq.~\ref{eqS:Dicke} and they form a so-called Decoherence-Free Subspace (DFS)\cite{zanardi97a,lidar98a}, and are therefore uncoupled from dissipation, whereas 
the 
states with $J=N/2,\ m_J >-N/2$ are super-radiant with an enhanced decay rate proportional to the atom number $N$. It is interesting to emphasize that we can find similar physics, i.e., pure Dicke model, using an atomic configuration where the $z_n = n \lambda_\aa/2=n \pi/q(\omega_\aa)$ ($n \in \N$). The difference between the two configurations is the symmetry of the super/subradiant states, which has to be taken into account in the generation of these states.

\subsection{Controlling atomic states under strong dissipation.}
Our first goal is to generate particular superpositions of atomic states within the DFS; therefore we need to include in our system some additional fields that allow to control the individual atomic states. We find convenient to introduce an extra auxiliary level $\ket{s}_n$, as sketched in Fig. \ref{fig1}(b) of the main text and use the following fields to control the atomic state:
\begin{align} 
H_{\mathrm{las}}&=\left( \frac{\Omega_r}{2}\sum_{n=1}^{N} \sigma^n_{se} + \frac{\Omega_\mathrm{anc} }{2} \sigma^{N+1}_{se}+\mathrm{h.c.} \right) + \Delta_e \sum_{n=1}^{N+1} \sigma_{ee}^n\,, \label{eqS:laser} \\
H_{\mathrm{c}}&=\frac{\Omega_{\mathrm{c}}}{2} \sigma^{N+1}_{sg} + \mathrm{h.c.}\,,
\label{eqS:laserb}
\end{align}
where $H_{\mathrm{las}}$ allows to control both the coupling between atom and 1D-reservoir, while $H_{\mathrm{c}}$ allows to control the atomic state of the ancilla atom independently of the coupling to the reservoir. We are interested in working in the regime of strong collective dissipation,
where $N\Gamma_{\oned}\gg\Omega_r, \Omega_\mathrm{anc},\Omega_{c},\Delta_e$. 
In this situation, we can intuitively consider that the 1D bath is continuously \emph{monitoring} the atomic state, as in the Quantum Zeno regime \cite{zanardi97a,lidar98a,beige00a,facchi02a}, and projecting the atomic state into the DFS of the Liouvillian $\mathcal{L}_\DD$. Notice that when including the extra auxiliary level $\ket{s}$, the $\mathrm{DFS}\equiv \{\ket{\Psi}:\ S_{ge}\ket{\Psi}=0\}$ contains all superposition states of the metastable states $\ket{g}$ and $\ket{s}$ plus all the excited states from the nullspace of the collective dipole $S_{ge}$. Formally, we obtain effective dynamics within the DFS by using a projector operator $\mathbb{P}$ satisfying: $\mathbb{P}{\cal L}_{\rm D}={\cal L}_{\rm D} \mathbb{P}=0$, and its orthogonal part: $\mathbb{Q}=1-\mathbb{P}$. Using these 
projectors, one can formally integrate out the fast dynamics outside of the DFS, 
described by $\QQ \rho$, and obtain effective dynamics of the atomic system within the DFS given by \cite{gardiner_book00a}
\begin{equation}
 \PP \dot{\rho}=\PP W \PP \rho-\PP W \QQ \frac{1}{{\cal L}_{\rm D}}\QQ W \PP \rho+\mathrm{O}[\frac{\tau^-3}{\Gamma_{\oned}^2}]\,.
\end{equation}
where $W$ is any perturbation acting on the atomic system (with relevant time scale $\tau$), e.g., $W=H_{\mathrm{las}}+H_{\mathrm{c}}$ and $\tau\Gamma_\oned$ must be $\ll 1$ such that it is a good approximation.

\subsection{Preparation of many-body entangled states: neglecting losses.}

As explained in the main manuscript, the first step of our protocol consists of creating a certain class of states that must satisfy: i) they must be easily mapped to the superradiant states of the atomic ensemble; ii) they must be created using only states within the DFS such that we can avoid the dissipation induced by the waveguide. We propose in Fig.\ref{fig1}(a) of the main manuscript a configuration of $N$ atoms and a separately addressable ancilla atom. Because of the high symmetry of the states that we aim to create among the first $N$ atoms, it is convenient to introduce the following notation to describe any symmetric combination of states over these atoms:
\begin{align}
\ket{F_{m,k}}=\mathcal{N}(m,k)^{-1/2}\mathrm{sym}\{\ket{s}^{\otimes m}\otimes\ket{e}^{\otimes k}\otimes\ket{g}^{\otimes N-m-k}\}\,,
\end{align}
where $\mathcal{N}(m,k) = \binom {N} {m,k,N-m-k}$ is the multinomial coefficient that gives the normalization of these states. This notation embeds the many-body entangled states that we aim to create, i.e., $ \ket{F_{m,0}}\equiv \ket{D_m}$ and $ \ket{F_{0,m}}\equiv \ket{S_m}$. For the ancilla atom, we use a notation $\ket{\psi}_\mathrm{A}$. 

\begin{figure}
	\centering
	\includegraphics[width=0.98\textwidth]{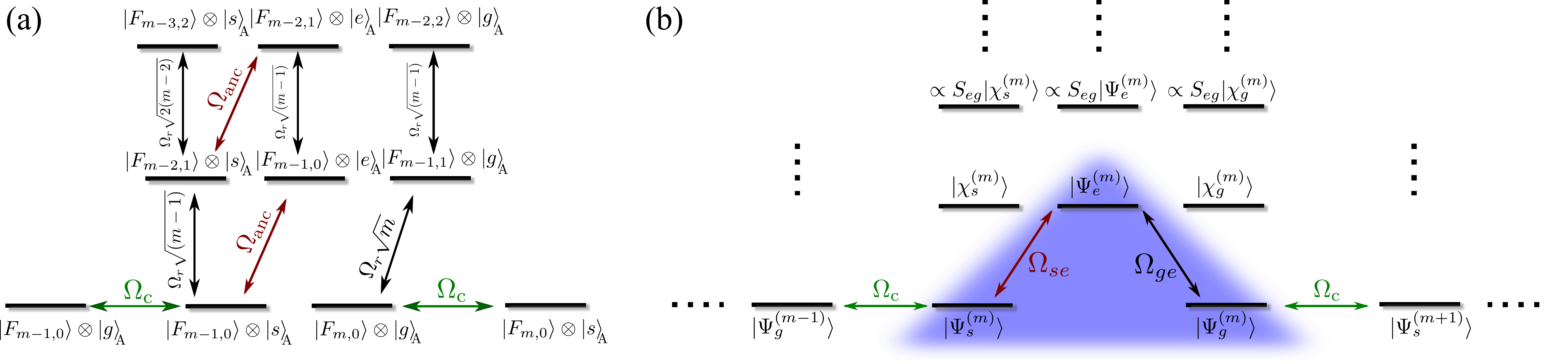}
	\caption{(a) Product of the Hilbert space of $N$ permutation invariant emitters and of one ancilla for $m$ excitations, that is $m$ emitters in state $\ket{s}$ or $\ket{e}$. (b) Separation of whole Hilbert space (for $m$ excitations) into DFS states (blue background) and non-DFS states, which makes obvious the emergence of the effective $\Lambda$-type transitions within the DFS.}\label{fig4}
\end{figure}

In Eqs. \ref{eqS:laser}-\ref{eqS:laserb} we introduced the laser configuration that we use, namely, a symmetric excitation over the first $N$ atoms and a different one for the ancilla. Interestingly, the combination of this configuration and the collective dissipation imposes certain symmetry conditions on the states that can exist within the DFS, namely, the atomic states with excited states ($\ket{e}$) must be symmetric under any permutation of the first $N$ atoms and antisymmetric with the permutation of the ancilla. This reduces the exponential Hilbert space of all states ($3^m$) to a set of $5m$ relevant states, which are depicted in Fig.~\ref{fig4}. For example, among all the combination of states $\ket{F_{m,1}}$ with one atomic excited state, for each $m$ only one of them (denoted by $\ket{\Psi_e^{(m)}}$) belongs to the DFS. Moreover, the states of combinations of the ancilla with $\ket{F_{m,2}}$ are all superradiant as none of them can fulfil the symmetry to be within the DFS of the collective 
dipole $S_{eg}$.

Now, let us explain how to generate the $\ket{D_m}$ using the tools that we have introduced. As shown in Fig. \ref{fig4}(b), for each $m$,  there exists an effective $\Lambda$-scheme within the DFS of the Liouvillian $\mathcal{L}_D$ that couples the states $\ket{\Psi_{s}^{(m)}}=\ket{D_{m-1}}\otimes\ket{s}_{A}$ to $\ket{\Psi_{g}^{(m)}}=\ket{D_m}\otimes 
\ket{g}_{A}$, through an excited state $\ket{\Psi_{e}^{(m)}}$ that also belongs to the DFS. If no projection into the DFS is considered, the states $\ket{\Psi_{g,s}^{(m)}}$ are coupled with excited states as follows:
\begin{align}
\label{eqs:las}
 H_{\mathrm{las}}\ket{\Psi_{s}^{(m)}}&= \frac{\Omega_{r}}{2} \sqrt{m-1} \ket{F_{m-2,1}}\otimes\ket{s}_\mathrm{A}+ \frac{\Omega_{\mathrm{anc}}}{2}\ket{F_{m-1,0}}\otimes\ket{e}_\mathrm{A}\,,\\
 H_{\mathrm{las}}\ket{\Psi_{g}^{(m)}}&= \frac{\Omega_{r}}{2}\sqrt{m}\ket{F_{m-1,1}}\otimes\ket{g}_\mathrm{A}\,,\nonumber 
\end{align}

Interestingly, it is possible to write Eqs.~\ref{eqs:las} separating the contributions of the states in and out of the DFS:
\begin{align}
\label{eqs:las2}
 H_{\mathrm{las}}\ket{\Psi_{s}^{(m)}}&= -\frac{\Omega_{\mathrm{anc}}}{2}\sqrt{\frac{\Nm}{\Nm+1}}\ket{\Psi_{e}^{(m)}}+\frac{\Omega_{\mathrm{anc}}}{2}\sqrt{\frac{1}{\Nm+1}}\ket{\chi_g^{(m)}}+\frac{\Omega_{r}}{2}\sqrt{m-1} \ket{\chi_s^{(m)}}\,, \\
   H_{\mathrm{las}}\ket{\Psi_{g}^{(m)}}&= \sqrt{\frac{m}{\Nm+1}}\frac{\Omega_{r}}{2}\ket{\Psi_e^{(m)}} +\frac{\sqrt{m \Nm}}{\Nm+1}\frac{\Omega_{r}}{2}\ket{\chi_g^{(m)}}\,,\nonumber
\end{align}
where we have introduced the notation $\Nm=N-m+1$, and $\ket{\Psi_e^{(m)}}$ is a state within the DFS that couples both $\ket{\Psi_{s,g}^{(m)}}$, and is given by:
\begin{align}
 \ket{\Psi_e^{(m)}}&=\sqrt{\frac{\Nm}{\Nm+1}}\ket{F_{m-1,0}}\otimes\ket{e}_\mathrm{A}\nonumber -\frac{1}{\sqrt{\Nm+1}}\ket{F_{m-1,1}}\otimes\ket{g}_\mathrm{A}\,.
\end{align}
$\ket{\chi_{s/g}^{(m)}}$ are two states outside the DFS defined as
\begin{align}
 \label{eq:super}
 \ket{\chi_s^{(m)}}&=\ket{F_{m-2,1}}\otimes\ket{s}_\mathrm{A}\,,\\ \nonumber
 \ket{\chi_g^{(m)}}&=\frac{1}{\sqrt{\Nm+1}}\ket{F_{m-1,0}}\otimes\ket{e}_\mathrm{A}+\sqrt{\frac{\Nm}{\Nm+1}}\ket{F_{m-1,1}}\otimes\ket{g}_\mathrm{A}\,,\nonumber
\end{align}
and can be shown to have an enhanced decay rate of $\Gamma_e=(\Nm+1)\Gamma_{\oned}$, by looking at the action of the collective operator $S_{eg}$ on these states. We discuss first the effect considering only perturbations up to first order within the Zeno dynamics. Therefore, the super-radiant states can be neglected as they are only virtually populated due to their enhanced decay rate. Thus, we first consider the effective $\Lambda$ system, with effective Raman intensities given by (see Fig. \ref{fig4}(b)):
\begin{align}
 \Omega_{se}^{(m)} & = \bra{\Psi_e^{(m)}} H_\mathrm{las} \ket{\Psi_s^{(m)}} =-\frac{\Omega_{\mathrm{anc}}}{2}\sqrt{\frac{\Nm}{\Nm+1}}\,,\\\,
 \Omega_{ge}^{(m)} & = \bra{\Psi_e^{(m)}} H_\mathrm{las} \ket{\Psi_g^{(m)}} =\frac{\Omega_{r}}{2}\sqrt{\frac{m}{\Nm+1}}\,. \nonumber
\end{align}
 where we see the importance of addressing the ancilla atom individually from the other $N$ emitters and keeping $\Omega_r\neq \Omega_{\mathrm{anc}}$, as we can now set them such that $|\Omega_{se}^{(m)}|=|\Omega_{ge}^{(m)}|$, by choosing: $| \Omega_{\mathrm{anc}}| =|\Omega_{r}| \sqrt{m/\Nm}$. This choice allows us to compensate the different Stark-shifts that are introduced with the projection $\PP$ and yields an off-resonant two-photon transition with Rabi frequency: 
\begin{align}
 |\Omega^{(m)}| =\frac{|\Omega_r|^2}{2\Delta_e} \frac{m}{\Nm+1}\,.
\end{align}

Notice then, that by flipping the state of the ancilla with $\Omega_{\mathrm{c}}$, one also flips $\ket{\Psi_{g}^{(m)}}\rightarrow \ket{\Psi_s^{(m+1)}}$, which re-initializes the process. Thus, by using a combination of $m$ off-resonance Raman transition and $m$ control fields, we can generate any $\ket{D_m}$ (or superpositions thereof).

\subsection{Preparation of many-body entangled states: error analysis.\label{subsec:error}}

By using the effective hamiltonian $H_{\mathrm{eff}}=\mathcal{P}(H_{\mathrm{las}}+H_{\mathrm{c}})\mathcal{P}$ (with $\mathcal{P}$ being the hamiltonian projection into the DFS) under the conditions described in the previous Section, we see that the time of the operation to do a complete transfer of population from $\ket{\Psi_{s}^{(m)}}$ to $\ket{\Psi_{g}^{(m)}}$ by using an off-resonant transition, i.e.,  $\Delta_e\gg \Omega_{ge,(se)}^{(m)}$, is given by:
\begin{equation}
 t_{\mathrm{op}}^{(m)}=\frac{\pi }{|\Omega^{(m)}|}\approx \frac{2\pi \Delta_e N}{m|\Omega_r|^2}\,.
\end{equation}

So far, we have considered the ideal situation; however, to estimate the fidelities in the preparation of these states, we need to analyze the errors that may occur within $t_{\mathrm{op}}^{(m)}$. The errors come from:
\begin{itemize}
 \item The spontaneously emitted photons from $\ket{\Psi_e^{(m)}}$ to other decay channels other than the waveguide, described by a Lindblad term: ${\cal L}_{*}(\rho) = 
\sum_{n} \frac{\Gamma^*}{2} \left( \sigma^n_{ge} \rho \sigma^n_{eg} - \rho \sigma^n_{ee}  \right)
+ \rm {H.c.}\, $, that we embedded in a single decay rate $\Gamma^*$. As we are using an off-resonant Raman transition, this source of error scales (for $N\gg m \ge 1$):
 \begin{equation}
  \epsilon_{\Psi_e}^{(m)}=\Gamma^*\frac{m|\Omega_r|^2}{4(\Nm+1)(\Delta_{e}^2+(\Gamma^*)^2)}\approx \Gamma^*\frac{m|\Omega_r|^2}{4 N\Delta_{e}^2}\,,
  \end{equation}
 where we used that $\Delta_e\gg \Gamma^*$.

 \item The other errors may appear due to photons emitted from the states out of the DFS, i.e., through the  $\ket{F_{m,1}}$-like states.  We can estimate the rate of these errors  which are finally given by (for $N\gg m \ge 1$):

  \begin{align}
    \label{eqS:erroff}
     &\epsilon_{\chi_s^{(m)}}=\big(\Gamma^*+N\Gamma_{\oned}\big)\frac{m|\Omega_r|^2}{4N^2\big(\Delta_{e}^2+(\Gamma^*+N\Gamma_\oned)^2\big)}\,,\\
  &\epsilon_{\chi_g^{(m)}}=\big(\Gamma^*+N \Gamma_{\oned}\big)\frac{m|\Omega_r|^2}{4\big(\Delta_{e}^2+(\Gamma^*+N \Gamma_{\oned})^2\big)}+\big(\Gamma^*+N\Gamma_{\oned}\big)\frac{m|\Omega_r|^2}{4N\big(\Delta_{e}^2+(\Gamma^*+N\Gamma_{\oned})^2\big)\big)}\,, \label{eqS:erroff1}
  \end{align}
  
\end{itemize}

Using these estimations, and with the following hierarchy for the parameters: $N\gg m\ge1$ and  $\Gamma_{\oned}\gg\Delta_e\gg \Gamma^*$ to simplify the expressions, we find the infidelity of the step $m$ to be
\begin{equation}
1-F^m=t_{\mathrm{op}}^{(m)}\Big(\epsilon_{\Psi_{e}}^{(m)}+\epsilon_{\chi_s^{(m)}}+\epsilon_{\chi_g^{(m)}}\Big)\approx \frac{\pi }{2}\big(\frac{\Gamma^*}{\Delta_e}+\frac{\Delta_e}{\Gamma_{\oned}}\big)\,,
\end{equation}
which is optimized for $\Delta_{e,\mathrm{opt}}=\sqrt{\Gamma^* \Gamma_{\oned}}$ yielding a scaling $1-F^m_{\mathrm{opt}}\propto 1/\sqrt{P_{\oned}}$. Interestingly, the scaling does not depend on either the number of atoms, $N$, nor the number of excitations, $m$. 

So far, we have focused our discussion on the $m$-th step that goes from $\ket{D_{m-1}}\rightarrow \ket{D_{m}}$. As depicted in Fig.~\ref{fig2} of the main manuscript, by combining this process with a transition over the ancilla qubit that initializes it in $\ket{s}_A$, we can generate any arbitrary superposition of states (over the first $N$ atoms) up to a $m_{\mathrm{max}}$, namely $\ket{\Psi_{D}}=\sum_{m=0}^{m_{\mathrm{max}}} d_m \ket{D_{m}}$, with a total of $m_{\mathrm{max}}$ Raman transitions together with $m_{\mathrm{max}}$ initialization gates on the ancilla. Neglecting the errors of the microwave transition over the ancilla, the total infidelity to generate $\ket{\Psi_{D}}$ is:
\begin{equation}
1-F_{\mathrm{opt}}\propto \frac{m_{\mathrm{max}}}{\sqrt{P_{\oned}}}\,.
\end{equation}

\subsection{Preparation of many-body entangled states: numerical analysis.}

In order to validate our scaling analysis, we study the preparation of two relevant sets of states without doing any approximation and considering all the possible states (including super-radiant ones). The two sets of states are i) the general class of states $\ket{D_m}$ and ii) the superpositions $\ket{\Phi_m}=\frac{1}{\sqrt{2}}\big(\ket{D_0}+\ket{D_m}\big)$.

First, it is interesting to realize that the symmetry conditions found from our analysis of the ideal situation tell us the relevant Hilbert space of the problem can be written in terms of the states: $\ket{F_{m,k}}\otimes \ket{\psi}_\mathrm{A}$, where $\ket{\psi}_\mathrm{A}=\{\ket{g}_\mathrm{A},\ket{s}_\mathrm{A},\ket{e}_\mathrm{A}\}$ is the Hilbert space of the ancilla atom, $k$ can be restricted up to 2 photons (as higher excited states will be only weakly populated), and $m=0,1,\dots, \mmax$, where $\mmax$ is the highest excitation that we want to achieve. Notice, that the Hilbert space does not depend directly on the total atom number $N$, but only on $\mmax$. The number of atoms, $N$, only enters in the two-photon resonant condition that fixes $\Omega^{(m)}$. In this Hilbert space we can use a non-Hermitian evolution, where the effective Hamiltonian is determined by the action of $H_{\mathrm{las}}+H_{c}$ plus the imaginary energies determined by the coupling to the 
waveguide through the $S_{ge}$ operator. The non-Hermitian Hamiltonian elements are given by:
\begin{align}
 \label{eqS:nhham}
 &\bra{F_{n,q}}\otimes  \bra{\phi}_\mathrm{A} (H_{\mathrm{las}}+H_{c})\ket{F_{m,k}}\otimes \ket{\psi}_\mathrm{A}= k[\Delta_e-i((N-m-k+1)\Gamma_\oned/2 +\Gamma^*/2)]\delta_{k,q}\delta_{m,n}\delta_{\psi,\phi}+\nonumber \\
 &+ [ \frac{\Omega_r}{2} \sqrt{m(k+1)} \delta_{k+1,q}\delta_{m-1,n}\delta_{\psi,\phi}+\mathrm{H.c.}]+ [\frac{\Omega_\mathrm{anc}}{2} \delta_{k,q}\delta_{m,n}\delta_{\psi,s}\delta_{\phi,e}+ \mathrm{H.c.}]+ [\frac{\Omega_\mathrm{c}}{2} \delta_{k,q}\delta_{m,n}\delta_{\psi,s}\delta_{\phi,g}+ \mathrm{H.c.}]\,.
\end{align}

A diagram of the relevant transitions in the complete Hilbert space is depicted in Fig.~\ref{fig4}(a). For the generation of individual Fock states, the pulse sequence can be easily deduced. One just needs to ensure a complete transfer of populations from $\ket{F_{0,0}}\otimes\ket{g}_\mathrm{A}\rightarrow \ket{F_{0,0}}\otimes\ket{s}_\mathrm{A} \rightarrow \ket{F_{1,0}}\otimes\ket{g}_\mathrm{A} \rightarrow \ket{F_{1,0}}\otimes\ket{s}_\mathrm{A}\rightarrow \ket{F_{2,0}}\otimes\ket{g}_\mathrm{A}\dots \rightarrow \ket{F_{m,0}}\otimes\ket{g}_\mathrm{A}$, which can be done by fixing the time of interaction, $t$, to $t\Omega_{\mathrm{c}}=\pi$ ($t\Omega^{(m)}=\pi$) for the microwave (two-photon Raman) transitions.
In Fig.~\ref{fig3}(a) of the main text, we show the numerical fidelities obtaining when fixing the off-resonant transition to the optimal $\Delta_{e,\mathrm{opt}}$ that we explored in the previous Section: the dots correspond to the numerical fidelities, whereas the solid lines depict the scaling $\propto 1/\sqrt{P_{\oned}}$ and show how our general arguments give us the right scaling. For a more complicated state, such as the $\ket{\Phi_m}$ the pulse sequence can be calculated numerically. In Fig.~\ref{fig3}(b), we show the optimal fidelities for generating these states up to 5 excitations, showing again how the $1/\sqrt{P_\oned}$ scaling of fidelities also holds for superpositions.

\subsection{Conditional preparation of many-body entangled states: using post-selection.}

In the error analysis we made in Section~\ref{subsec:error}, we realized that some of the errors were coming from the small populations of superradiant states $\ket{\chi_{g,s}^{(m)}}$ that emit quickly into the waveguide. Actually, in our atom-waveguide configuration one can think of using another atomic ensemble that acts as an efficient photonic absorber that maps the photonic excitation into an collective atomic one that afterwards can be detected via fluorescence with a very high fidelity. Moreover, it is possible to use a more elaborate scheme as sketched in Ref. \cite{chang12a}, where collective atomic excitations can be mapped to a single impurity atom, making fluorescence detection much more efficient. As it is not the purpose of our manuscript to elaborate on conditional preparation, we just assume that we can perfectly detect all the photons emitted through the waveguide and study the scaling of the fidelities. Using these assumptions, we can cancel the errors in Eqs.~\ref{eqS:erroff} and~\ref{eqS:erroff1} that are 
proportional to $N\Gamma_{\oned}$, that in the limit $N\gg m\ge1$ yield:

\begin{align}
    \label{eqS:erroffpost}
     &\epsilon_{\chi_s^{(m)}}\approx\Gamma^*\frac{m|\Omega_r|^2}{4N^2\big(\Delta_{e}^2+(\Gamma^*+N\Gamma_\oned)^2\big)}\,,\\
  &\epsilon_{\chi_g^{(m)}}\approx\Gamma^*\frac{m|\Omega_r|^2}{4\big(\Delta_{e}^2+(\Gamma^*+N \Gamma_{\oned})^2\big)}+\Gamma^*\frac{m|\Omega_r|^2}{4N\big(\Delta_{e}^2+(\Gamma^*+N\Gamma_{\oned})^2\big)\big)}\,.
  \end{align}

We can immediately see that the leading error comes from the first contribution of $\epsilon_{\chi_g^{(m)}}$. Taking the leading error only, we can then estimate the infidelity of the step $m$ to be
\begin{equation}
1-F^m=t_{\mathrm{op}}^{(m)}\Big(\epsilon_{\Psi_{e}}^{(m)}+\epsilon_{\chi_s^{(m)}}+\epsilon_{\chi_g^{(m)}}\Big)\approx \frac{\pi N \Gamma^*}{2}\Big(\frac{1}{\Delta_e}+\frac{\Delta_e}{(N\Gamma_{\oned})^2}\Big)\,,
\end{equation}
which is optimized for $\Delta_{e,\mathrm{opt}}=N\Gamma_{\oned}$ yielding a scaling $1-F^m_{\mathrm{opt}}\propto 1/P_{\oned}$. It is important to highlight that the optimal condition can not be realized as it implies $\Delta_{e,\mathrm{opt}}\gg \Gamma_{\oned}$, which violates the conditions under which we derived our effective Hamiltonian. However, the linear improvement with $1/P_{\oned}$ is still obtained even if we do not reach the optimal conditions. More details, on how to take advantage of the atom nanophotonic waveguide for conditional preparation will be presented elsewhere.\cite{workinprogress15}

\section{Atom-photon mapping starting from an initial atomic excitations.}

Let us review first the general derivation for a Hamiltonian of the form $H = H_\mathrm{S} + H_\mathrm{B} + H_\mathrm{SB}$, with
\begin{align}
H_\mathrm{SB} =& \sum_{n,q} g_q a_q^\dagger O^n \ee^{-\ii q z_n}  + \mathrm{H.c.}\,,
\end{align}
where $H_S$ ($H_B$) are the system (1D-bath) Hamiltonians, and $H_\mathrm{SB}$ is the interaction between them, where $O^n$ ($a_q$) are the system (1D-bath) operators. Using the generalized input-output formalism \cite{caneva15a}, everything boils down to calculate the scattering amplitude
	\begin{align}
		A(t) = \bra{\phi_\mathrm{out}} \bra{B_\mathrm{out}} \ee^{-\ii H t } \ket{B_\mathrm{in}} \ket{\phi_\mathrm{in}},
	\end{align}
where $\ket{\phi_\mathrm{in(out)}}=\gamma_\mathrm{in(out)}^\dagger(t)\ket{\mathrm{vac}}$ denotes the system input (output) state at time $t$ and $\ket{B_\mathrm{in(out)}}$ the input (output) state of the bath, which in our case is the electromagnetic field inside the waveguide. Let us particularize to our situation of interest, where we decay from an initial system state with $m$ excitations, that is, $\gamma_\mathrm{in}^\dagger (0) \ket{\mathrm{vac}}=\ket{S_m}$, $\gamma_\mathrm{out}(T) = \mathbf{1}$ and $\mathcal{F}_\mathrm{in} = \mathbf{1}$ and the operators $O^n=\sigma_{ge}^n$. Moreover, we also linearize the waveguide dispersion relationship, i.e., $\omega_q\approx v_g(q_\aa) |q|$, assume for simplicity that: $g_q\approx g_{q_\aa}\equiv g$. With these considerations, the scattering amplitude for $m$ excitations simplifies to \cite{caneva15a,shi15a}
	\begin{align}
		& A_{ \{ q\}} (t) = \left( -\ii \right)^m g^m \int_{0}^{t} \dd s_1 \cdots  \int_{0}^{t}  \dd s_m\  \ee^{-\ii \sum_{i=1}^{m} \omega_{q_i} (T-s_i)} \times \bra{\mathrm{vac}} \mathcal{T} O_{q_1}(s_1)  O_{q_2}(s_2)  \cdots O_{q_m}(s_{m}) \ket{S_m}\,, 
	\end{align}
where $\{q\}=\{q_1,\dots,q_m\}$ is the set of relevant momenta of the $m$-photon state, where each of them run over the whole Brillouin Zone $q_i\in\mathrm{B.Z.}$ and $\mathcal{T}$ is the time ordering operator that guarantee that $s_1>s_2>\dots>s_m$. A further simplification is obtained if we assume that we are within the Markov approximation, and use the fact that we work with an atomic configuration such that: $q_\aa z_n=2\pi n$. Then,
	\begin{align}
		O_q =  \sum_n \sigma^n_{ge} \ee^{-\ii q z_n} \approx \sum_n \sigma^n_{ge} \ee^{-\ii q_\aa z_n} = S_{ge}\,.\label{eqS:approxO}
	\end{align}

With this approximation the time ordering simply ensures that the final (bosonic) state is symmetrized over all sets $\{ q \}$. Notice, that the output photonic state associated to this scattering amplitude can be written as

\begin{equation}
\ket{\Psi_{B}^{(m)}}=\sum_{\{ q\}} \frac{A_{ \{ q\}} (t)}{m!} \ud{a}_{q_1} \ud{a}_{q_2} \dots \ud{a}_{q_m}\ket{\mathrm{vac}}\,,
\end{equation} 
where the sum over ${ \{ q\}}=\{q_1,\dots,q_m\}$ extends over all momenta. The state $\ket{\Psi_{B}^{(m)}}$ is normalized with the $1/m!$ factor as it cancels the $m!$ terms that appears from the permutations of all the $a_{q_i}$'s and the $m!$-factor of the scattering amplitude normalization in the whole ${ \{ q\}}$-space, i.e., $\sum_{ \{ q\}} | A_{ \{ q\}} (t)|^2=m!$ (notice the scattering amplitude $A_{ \{ q\}}$ is normalized to one only if: $\sum_{q_1>q_2>\dots> q_m} | A_{ \{ q\}} (t)|^2=1$).

Therefore, it is enough to calculate the contribution of one time ordering, e.g., $s_1>s_2>\dots >s_m$, and then sum up to all the permutations of $\{ q\}$.  As was shown in the previous sections, the effective (non-hermitian) system Hamiltonian is
	\begin{align}
		H_\mathrm{eff} = \omega_\mathrm{a} S_{ee} - \ii (\Gamma_\oned/2) S_{eg} S_{ge}.
	\end{align}
Interestingly, our initial state $\ket{S_m}$ is an eigenstate of this effective  Hamiltonian, and we can now calculate the action of the operator 
	\begin{align}
		S_{ge}(s) \ket{S_m} =\sqrt{N_m} \sqrt{m}\ \ee^{\left[ -\ii \omega_\aa -\Gamma_{\oned} (m N_m-(m-1)N_{m-1})/2 \right] s} \ket{S_{m-1}}
	\end{align}
and hence the correlator
	\begin{align}
		&\bra{\mathrm{vac}} S_{ge}(s_1)  S_{ge}(s_2)  \cdots S_{ge}(s_{m}) \gamma_\mathrm{in}^\dagger (0) \ket{\mathrm{vac}}= \prod_{r=1}^{m}  \sqrt{ r N_r} \exp\Big[\left[-\ii \omega_\aa  -\Gamma_\oned (r N_r-(r-1)N_{r-1})/2  \right] s_r \Big].
	\end{align}

When doing the integral, one needs also to take care of the particular time ordering considered. In the case that we have chosen, $s_1>\dots>s_m$, the integral can be rearranged $\int_0^t \dd s_m \int_{s_{m}}^{t} \dd s_{m-1} \dots   \int_{s_2}^t \dd s_{1}$. The choice of the upper limit of integration to $t$ is not casual, as in each time integral we will obtain a term proportional to $\propto e^{- N \Gamma_{\oned}t/2 }$, that will disappear for times $t\gg 1/(N\Gamma_\oned)$ which are the ones that we are interested in. For example, the first integral:
\begin{align}
\int_{s_2}^t \dd s_1 \exp\Big[\left[\ii (\omega_{q_1}- \omega_\aa)  -\Gamma_\oned N_1/2  \right] s_1 \Big]&=-\frac{1}{\ii (\omega_{q_1}- \omega_\aa)  -\Gamma_\oned  N_1 /2 } \Big( \exp\Big[\left(\ii (\omega_{q_1}- \omega_\aa)-\Gamma_\oned N_1/2   \right) s_2 \Big]-\\ \nonumber &\exp\Big[\left(\ii (\omega_{q_1}- \omega_\aa) -\Gamma_\oned N_1 /2  \right) t \Big]\Big)\\ \nonumber
&\approx -\frac{1}{\ii (\omega_{q_1}- \omega_\aa)  -\Gamma_\oned N_1/2 } \,\exp\Big[\left(-\ii (\omega_{q_1}- \omega_\aa)  -\Gamma_\oned N_1/2   \right) s_2 \Big]\,,
\end{align}
where the last approximation was done for  $t\gg 1/(N\Gamma_\oned)$. The second integral then reads:
\begin{align}
-&\int_{s_3}^t \dd s_2 \frac{\exp\left[\left[\ii (\omega_{q_1}+\omega_{q_2}- 2\omega_\aa)  -\Gamma_\oned  2 N_2/2   \right] s_2 \right]}{\ii (\omega_{q_1}- \omega_\aa)  -\Gamma_\oned N_1/2 }=\\ \nonumber 
&\approx \frac{1}{[\ii (\omega_{q_1}- \omega_\aa)  -\Gamma_\oned N_1/2 ][\ii(\omega_{q_1}+\omega_{q_2}- 2\omega_\aa)- \Gamma_\oned  2 N_2/2 ]}\,\exp\Big[\left[\ii (\omega_{q_1}+\omega_{q_2}- 2\omega_\aa)  -\Gamma_\oned  2 N_2 /2  \right] s_3 \Big]\,,
\end{align}

Iterating this integration and considering the permutations of the $\{q\}$ due to the different time-orderings, we obtain the following expression for the scattering amplitude
\begin{align}
		A_{ \{ q \} }(t) = \ii^m g^m  \prod_{r=1}^{m} \frac{\sqrt{r N_r}\ \ee^{-\ii \omega_{q_r} t} }{ \ii ( \sum_{l=1}^r \omega_{q_l}-r \omega_\aa) + r \Gamma_\oned N_r/2 } +[\{q\}-\mathrm{permutations}]\,
		\label{eqS:amplitude}
\end{align}
for sufficiently large times $t \gg 1/ N_m \Gamma_\oned$, that is when the system state has completely decayed and all the excitations have been transferred to the bath state. Notice that the only dependence on $t$ in this case enters through: $\ee^{-\ii \sum_{r=1}^m \omega_{q_r} t} $, which describes the center of mass motion of the wavepacket when going to the real space. In the low excitation regime, one can either do a Holstein-Primakoff approximation \cite{porras08a} or change $N_m\rightarrow N$, in the expression of Eq. \ref{eqS:amplitude}. In both cases we obtain:
\begin{align}
		A^{\mathrm{HP}}_{ \{ q \} }(t) =  \sqrt{m!} \ee^{-\ii \sum_{r=1}^m \omega_{q_r} t} \prod_{r=1}^{m} \frac{\ii g \sqrt{N} }{ \ii ( \omega_{q_r}- \omega_\aa) + \Gamma_\oned N/2 } =\sqrt{m!} \ee^{-\ii \sum_{r=1}^m \omega_{q_r} t} \prod_{r=1}^{m} C_{\Gamma_{\oned} N}(q_r) \,,
\end{align}
which represents a single mode wavepacket with spectral shape $C_{\Gamma_{\oned} N}(q)$. To emphasize the connection between the linear and non-linear scattering amplitudes we exemplify the results for the $m=2$ photon wavepacket. The non-linear scattering amplitude is given in this case by

\begin{align}
A_{ q_1,q_2 }(t) &=-g^2 \ee^{-\ii \sum_{r=1}^2 \omega_{q_r} t} \sqrt{2 N_2 N_1} \frac{1}{\ii(\omega_{q_1}+\omega_{q_2}- 2\omega_\aa)- \Gamma_\oned  2 N_2/2 }\Big[\frac{1}{\ii (\omega_{q_1}- \omega_\aa)  -\Gamma_\oned N_1/2 }+\frac{1}{\ii (\omega_{q_2}- \omega_\aa)  -\Gamma_\oned N_1/2 }\Big]=\nonumber\\
&=-g^2 \sqrt{2 N_2 N_1} \ee^{-\ii \sum_{r=1}^2 \omega_{q_r} t} \frac{1}{[\ii (\omega_{q_2}- \omega_\aa)  -\Gamma_\oned N_1/2 ][\ii (\omega_{q_1}- \omega_\aa)  -\Gamma_\oned N_1/2 ]}\Big[1+\Gamma_{\oned}\frac{(N_2-N_1)}{\ii(\omega_{q_1}+\omega_{q_2}- 2\omega_\aa)- \Gamma_\oned  2 N_2/2 }\Big]\,,
\end{align}
where one can clearly see how if $N_2=N_1$, then $A_{ q_1,q_2 }(t)\equiv A^{\mathrm{HP}}_{ q_1,q_2 }(t)$. Moreover, as $N_1-N_2=1$, then, it is direct to see that the correction due to the $N_2-N_1$ term is of the order $O(1/N)$. To generalize to higher photon numbers it is more convenient to directly calculate the overlap between the linear and non-linear approximations before doing the time integral. The reason is that one can formally integrate in $\{q\}$ variables before the times $s_i$, to obtain: 
\begin{equation}
\sum_{q_i} |g|^2 \exp\left[\ii (\omega_{q_i}  -\omega_{\aa})(s_r-\tilde{s}_r )\right]=\frac{2L|g|^2}{v_g(\omega_\aa)}\delta(s_r-\tilde{s}_r)=\Gamma_{\oned}\delta(s_r-\tilde{s}_r)\,.
\end{equation}

With these $\delta$'s the double time integral appearing when calculating the overlap is much simplified:
\begin{align}
\label{overlap}
\braket{\Psi_{B}^{(m)}}{\Psi_{B,\mathrm{HP}}^{(m)}}=\sum_{\{q\}}\frac{A^*_{\{q\}}(t)A^{\mathrm{HP}}_{\{q\}}(t)}{m!}= \Gamma_{\oned}^m \int_{0}^{t}\dd s_m  \cdots  \int_{s_2}^{t}  \dd s_1 \prod_{r=1}^{m}  r \sqrt{N N_r} \exp[-\Gamma_\oned (r N_r-(r-1)N_{r-1})/2  ]]\exp[-\Gamma_\oned r N/2  ]\,,
\end{align}
where in the last equality we have used the fact that the contribution of $m!$ different time-orders will be the same. Then, the multi-time integral can be calculated iteratively, yielding to an overlap:
\begin{align}
\label{overlap2}
1-\braket{\Psi_{B}^{(m)}}{\Psi_{B,\mathrm{HP}}^{(m)}}= 1-2^m\prod_{r=1}^{m}  \frac{\sqrt{N N_r}}{N+N_r}\approx  1-\prod_{r=1}^{m}  \frac{\sqrt{1-r/N}}{1-r/(2N)}\approx \frac{m^3}{20 N^2}+O(m^4/N^3)\,.
\end{align}
From the expression above, one can also check that $\ket{\Psi_{B}^{(m)}}$ is normalized by making $N_r\equiv N$. For consistency, one can also check that each $C_{\Gamma_{\oned} N}(q)$ in the linear expresion of $A_{q}^{\mathrm{HP}}(t)$ is normalized independently:
\begin{align}
\sum_q  |C_{\Gamma_{\oned} N}(q)|^2 &=\int_{-\infty}^\infty \frac{\dd q |g|^2 N L}{2 \pi}  \frac{1}{ (\omega_{q}- \omega_\aa)^2 +(\Gamma_\oned N/2 )^2} \nonumber \\
&\approx \int_0^\infty \frac{\dd \omega  |g|^2 N L}{v_g(\omega_\aa)  \pi}  \frac{N}{ (\omega_{q}- \omega_\aa)^2 +(\Gamma_\oned N/2 )^2}=\frac{ L |g|^2 N v_g(\omega_\aa)}{\pi} \frac{2\pi}{\Gamma_{\oned} N} =1
\end{align}

\section{Implementation details: Photonic Crystal Waveguides.}

A particularly promising system to implement our proposal is atom 1D nanophotonic systems, in which first working proof-of-principles examples have been realized by using ``alligator'' photonic crystal waveguides \cite{goban13a,goban15a}. In these systems, the renormalized spontaneous decay rate is given by:

\begin{equation}
 \frac{\Gamma_{\oned}}{\Gamma_\aa}=\frac{n_g \sigma \xi}{2 A_m}\,,
\end{equation}
where $n_g=c/v_g$ is the group index, $\sigma=3 \lambda_0^2/(2\pi)$ the radiative cross-section, $A_m$ the effective mode area and $\xi$ is an adimensional factor of cavity enhancement due the reflections at the end of the dielectric waveguide. Current SiN structures \cite{goban13a,goban15a}, have $A_m\approx 0.2$ $\mu$m$^2$,  $n_g\approx 10$ and cavity enhancement $\xi\sim 5$. There are several sources of errors in these systems:
\begin{enumerate}
 \item Spontaneous emission to other modes different from the chosen guided mode. Current structures show $\Gamma^*\sim \Gamma_a$, however, further design may result in further reduction of spontaneous emission, e.g., by making thicker dielectric structures $\Gamma^*\sim 0.1\Gamma_\aa$ \cite{gonzaleztudela14c}. Depending on the reduction of spontaneous emission $\Gamma^*=\alpha\Gamma_{\aa}$, the Purcell factor with current designs can be $P_\oned\sim 50/\alpha$.
 
 \item Intrinsic losses of the material yield finite $Q$-factors, which can be calculated as:
 \begin{equation}
  Q=\frac{n_r}{2n_i}\,
 \end{equation}
 with $n=n_r-in_i$ the refractive index of the material. The $Q$-factor can be easily related to the attenuation of the intensity of the field traveling through the dielectric as follows:
 \begin{equation}
  L_{\mathrm{prop}}\approx \frac{\lambda_0}{4\pi n_i n_g}\approx \frac{Q \lambda_\aa}{2\pi n_g}\,,
 \end{equation}
  with $\lambda_\aa=\lambda_0/n_r$ and where $L_{\mathrm{prop}}$ incorporates both material absorption via $n_i$ and effect of reduced group velocity. We notice that $Q$-factor also has contribution of scattering losses of the material due to imperfections, and therefore one must consider state-of-the-art values for doing estimations. For Cs atoms ($\lambda_0=894$ nm) and SiN structures ($n_r=2$, $Q\sim 10^6$, $n_g=10$), this yields $L_\mathrm{prop}/\lambda_\aa \gtrsim 10^4$. The main effect is that $J_{mn}$ must be corrected by this attenuation length: $J_{mn}\approx\Gamma_{\oned} e^{iq(\omega_\aa)|z_{mn}|} e^{-|z_{mn}|/L_{\mathrm{prop}}}$\cite{gonzaleztudela11a}, with $z_{mn}=z_m-z_n$. As $L_{\mathrm{prop}}/\lambda_\aa$ is very large, in our situation of interest with $z_n=n\lambda_\aa$, the effect of the finite propagation can be treated as a perturbation to the collective Liouvillian given by:
  \begin{equation}
    \label{eqS:mequation1}
      \mathcal{L}_{\mathrm{prop}}(\rho)= 	
	\sum_{n,m} \frac{\Gamma_{\oned}}{2}(1- e^{-|z_{mn}|/L_{\mathrm{prop}}})\left( \sigma_{ge}^n \rho \sigma_{eg}^m - \rho \sigma_{eg}^m \sigma_{ge}^n \right)
      + \rm {H.c.}\approx  	
	\sum_{n,m}\frac{\Gamma_{\oned}}{2}\frac{|z_{mn}|}{L_{\mathrm{prop}}}\left( \sigma_{ge}^n \rho \sigma_{eg}^m - \rho \sigma_{eg}^m \sigma_{ge}^n \right)
      + \rm {H.c.} \,.
   \end{equation}
 and introduces small corrections to the superradiant decay rate $\ket{S_m}$ and spontaneous emission rate of $\ket{\Psi_{e}^{(m)}}$ as long as the size of the atomic ensemble is $N\lambda_\aa \ll L_{\mathrm{prop}}$.
   
  \item If one thinks of increasing $\Gamma_{\oned}$ only through group velocity reduction another effect to take into account is retardation. The worst-case correction of this effect appears after doing a fast-resonant $\pi/2$ pulse to switch from $\ket{D_m}$ to $\ket{S_m}$ to do the atom-photon mapping. To observe superradiant behaviour in that case, it must be satisfied $N\Gamma_{\oned}<2 v_g/(N \lambda_\aa)$ (assuming $\lambda_\aa/2$ separation of atomic states), that with current state-of-art parameters \cite{goban13a,goban15a}, leads to $N\lesssim 500$. Notice, that in the preparation of superposition of $\ket{D_m}$, this critical number goes up to $N\lesssim 10^4$, as the characteristic timescales do not show the collective enhancement.
  
  Furthermore, there are several ways of overcoming retardation in these set-ups: i) increase $\Gamma_{\oned}$ not only by $v_g$ but through cavity enhancement, e.g., by placing mirrors at the end of dielectric \cite{goban15a}; ii) more easily by reducing the characteristic timescale doing the atom-photon mapping from $\ket{D_m}$ to $\ket{S_m}$ off-resonantly by setting a finite $\Delta_e\neq 0$. This reduces $\Gamma_{1d}$ (and $\Gamma^*$) by a factor $(\Omega_r/\Delta_e)^2$ which relaxes retardation requirements, while keeping $P_{\oned}$ constant.
  
  \item Moreover, a typical way of increasing group index is by using the regions of slow-light that appear close to 1D band-gaps, where one can approximate the dispersion relationship by $\omega_q\approx \omega_c+A(q-q_c)^2$. This dispersion will generate corrections with respect to the linear propagation of the wavepacket that must be kept small within the bandwidth $m N\Gamma_{\oned}$. 
\end{enumerate}

\bibliography{Sci,books}

\begin{thebibliography}{10}
\bibitem{footnote1}
\bibinfo{note}{Notice that with $d=n\lambda_{\mathrm{a}}/2$, similar physics
  can be observed if one adjusts the laser phases appropiately.}
\bibitem{footnote2}
\bibinfo{note}{Because part of the errors are coming from photons emitted into
  the waveguide, we can use another ensemble of atoms to absorb them
  efficiently and condition the preparation on detecting no atomic
  excitation~\cite{chang12a}. This leaves only the errors due to $\Gamma^*$,
  which leads to conditional fidelities of $1-F^{(m)}_{\mathrm{opt}}\propto
  \frac{1}{P_{\oned}}$ \cite{supmat}.}
\end{thebibliography}
\bibliographystyle{naturemag}
\end{document}